\documentclass{vldb}
\usepackage[latin1]{inputenc}
\usepackage{graphicx}
\usepackage{balance}  
\usepackage{color}
\usepackage{algorithmic}
\usepackage{algorithm2e}
\usepackage{tablefootnote}
\usepackage{etoolbox}
\usepackage{subcaption}
\usepackage{tikz}
\usetikzlibrary{circuits.logic.US, positioning} 
\usetikzlibrary{shapes.geometric, arrows}
\usepackage{mathpartir}
\usepackage{syntax}
\usepackage{newfloat}
\DeclareFloatingEnvironment[
  fileext   = logr,
  listname  = {List of Grammars},
  name      = Grammar,
  placement = htp
]{Grammar}

\newcommand{\cut}[1]{}
\newcommand{\sysname}{smcql\xspace}

\begin{document}


\title{{\Huge \sc \sysname}: Secure Querying for Federated Databases}



%
%
%
%

\numberofauthors{6} 

\author{
%
%
\alignauthor Johes Bater\\
\affaddr{Northwestern University}\\
\email{johes@u.northwestern.edu}
\alignauthor Gregory Elliott\\
\affaddr{Northwestern University}\\
\email{GregoryElliott2016@u.northwestern.edu}
\alignauthor Craig Eggen\\
\affaddr{Northwestern University}\\
\email{CraigEggen2016@u.northwestern.edu}
\and
\alignauthor Satyender Goel\\
\affaddr{Northwestern University}\\
\email{s-goel@northwestern.edu}
\alignauthor Abel Kho\\
\affaddr{Northwestern University}\\
\email{Abel.Kho@nm.org}
\alignauthor Jennie Rogers\\
\affaddr{Northwestern University}\\
\email{jennie@eecs.northwestern.edu}
}

\maketitle

\begin{abstract}

People and machines are collecting data at an unprecedented rate.    Despite this newfound abundance of data, progress has been slow in {\it sharing} it for open science, business,  and other data-intensive endeavors.  Many such efforts are stymied by privacy concerns and regulatory compliance issues.   For example, many hospitals are interested in pooling their medical records for research, but none may disclose arbitrary patient records to researchers or other healthcare providers.  In this context we propose the Private Data Network (PDN), a federated database for querying over the collective data of {\it mutually distrustful} parties.  In a PDN, each member database does not reveal its tuples to its peers nor to the query writer.   
Instead, the user submits a query to an honest broker that plans and coordinates its execution over multiple private databases using secure multiparty computation (SMC).  Here, each database's query execution is {\it oblivious}, and its  program counters and memory traces are agnostic to the inputs of others.

We introduce a framework for executing PDN queries\\ named {\sc \sysname}.   This system translates SQL statements into SMC primitives to compute query results over the union of its source databases without revealing sensitive information about individual  tuples to peer data providers or the honest broker.
Only the honest broker and the querier receive the results of a PDN query.  For fast,  secure query evaluation, we explore a heuristics-driven optimizer that minimizes the PDN's use of secure computation and partitions its query evaluation into scalable slices.  

\end{abstract}

\vspace{-5mm}
\section{Introduction}

Federated database systems, wherein many autonomous databases are united to appear as a single engine for querying, are having a renaissance in ``big data'' applications.   Interestingly, many such federations contain data owned by {\it mutually distrustful} parties who are willing to have the union of their data analyzed, but will not disclose their raw tuples.  We call a database federation that spans mutually distrustful sources  a {\it private data network} or PDN. Federations of this kind contain data that is privately held and not available for upload to the cloud.

In exploring this topic, we identified use cases for PDNs in medicine, data markets, banking, online advertising, and human rights work.   Typically, PDN members either upload their data to a trusted intermediary or they use one-off privacy-preserving algorithms  to mine it~\cite{Karr2005,Agrawal2000}.  We posit that PDNs will see broader adoption if their users express their analytics as declarative {\it SQL statements}.  

For example,  a consortium of hospitals is interested in pooling their patient records for clinical data research and each site is in charge of securing their own data.  A university researcher, operating independently of the hospitals, wants to evaluate a new treatment for rare disease $X$.  Her first step is to ask if there is a large enough cohort of $X$ sufferers in the consortium to form a study.   She writes {\tt SELECT COUNT(DISTINCT patient\_id) FROM diagnosis WHERE\\ diag=X;} to the consortium coordinator.  If this query were run in a standard database federation, the coordinator would collect and merge patient IDs for $X$ from each site, eliminate duplicates, and counts them.  This approach is undesirable because it reveals the patient IDs of individuals affected by $X$ to the coordinator.  A secure framework is needed that enables researchers to execute distributed queries without exposing information to unauthorized parties about their source data or intermediate results.

\begin{figure} [t]
\centering
\includegraphics[width=0.45\textwidth]{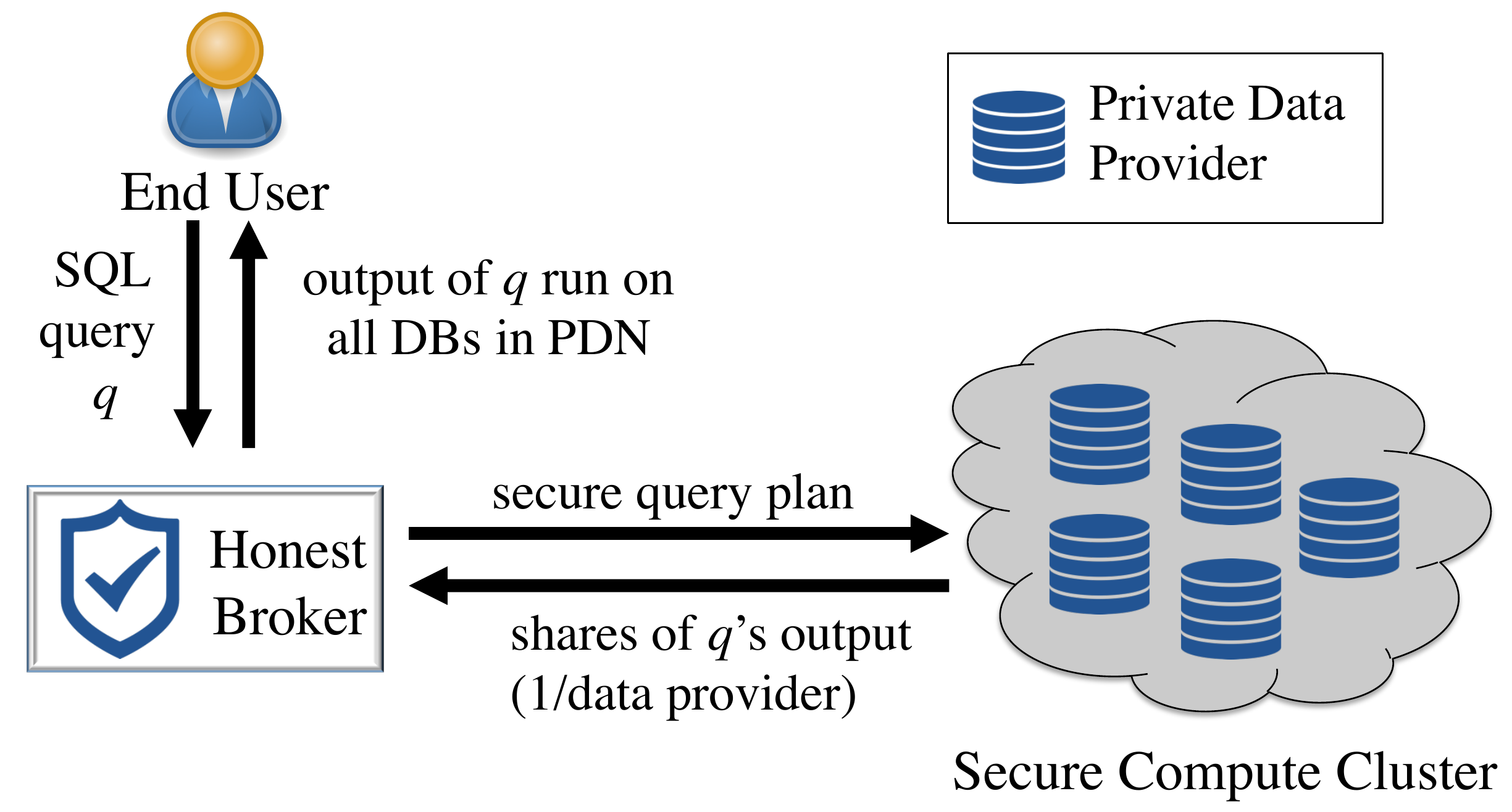} 
\caption{Private data network architecture}
\label{fig:conceptual-diagram}
\vspace{-6mm}
\end{figure}

The PDN architecture is shown in Figure~\ref{fig:conceptual-diagram}. Here, a user submits their query to the federation's {\it honest broker}, a neutral third party that plans and orchestrates its execution over two or more private data providers.  The federation has a shared schema that is supported by all parties. 
The execution of a PDN query is distributed over a secure compute cluster of {\it private data providers}.  Each provider executes a secure protocol provided by the honest broker that produces a share of the query output.  The honest broker assembles the shares into output tuples and sends them to the end user.   From the user's perspective, the PDN behaves exactly like a conventional federated database where one submits SQL and receives query results.  In our  example above, the end user is the researcher, the hospitals are private data providers, and the honest broker  is the consortium coordinator.



%
At setup time, the PDN has a shared set of table definitions. This schema is annotated with the level of protection required for each of its attributes.  For identifiers that span multiple data providers, such as patient IDs that appear in multiple hospitals, the honest broker works with the PDN members to carry out secure record linkage as in~\cite{al2005blocking, kho2015linkage,lazrig2015privacy}.

{\sc \sysname} is our framework for planning and executing PDN queries.  It uses secure multiparty computation (SMC) to evaluate distributed queries among mutually distrustful parties.   SMC is a subfield of cryptography that studies methods whereby two or more parties compute a function over their data while each keeps their individual inputs private.    SMC makes query evaluation  {\it oblivious} such that each party's computation is independent of the inputs of others.  An example of secure function evaluation is Yao's Millionaire Problem~\cite{Yao1982}. He asked, ``If Alice and Bob are millionaires who are interested in determining which one of them is richer, how can they solve for this without either party revealing their net worth?''.   Speaking imprecisely, SMC provides a black box within which the mutually distrustful parties combine their sensitive tuples for  query evaluation.   \cut{In this setting, neither party can learn  about the other's data by examining network transmissions, memory access traces, or program counters.  }

Since SMC computes distributed query evaluation obliviously,  it exacts a high performance cost in comparison to plaintext query evaluation.  For example, if one runs a join using SMC, the output of this operator will be the size of the cross product of its inputs.   Its output is padded with encrypted nulls to maintain the query's obliviousness.   With cascades of oblivious database operators, the cardinality of their intermediate results grows rapidly.  It is not uncommon to see secure functions run multiple orders of magnitude slower than their plaintext counterparts. 

In our work, a PDN seamlessly translates database operators within a query execution plan into SMC primitives.   It carefully manages its use of SMC so that its queries run efficiently.  The query planner first identifies when SMC is needed in a query by modeling the flow of sensitive data through its operator tree.  After that, it optimizes the subtree's execution using heuristics.  Lastly, the system generates secure code for the optimized plan.   In this paper, we present a query planner and executor for two mutually distrustful parties.

 \cut{Here, PDN queries use garbled circuits when they are evaluating steps in a query plan that compute on sensitive data.  }


{\sc \sysname} offers an honest-but-curious threat model.  In other words, we trust each data provider to faithfully execute the protocol provided by the honest broker.  On the other hand, participants may attempt to deduce the sensitive data of others during a secure query execution using side channels including program counters and memory accesses.  We do not address the question of malicious queriers who try to infer sensitive data by examining the output of many PDN queries.  Differential privacy~\cite{Dwork2006a}  tackles this issue by injecting controlled levels of noise into the data at query time to obscure the role of an individual in a larger dataset.  



In contrast, we propose a rule-based approach to protecting sensitive data that returns precise query results. Here, the PDN's stakeholders create a policy defining the queries they will admit for execution.  This policy may stipulate a minimum number of parties that must participate in each secure computation or attributes that are never accessible to end users in plaintext.  These rules reflect the cultural norms and best practices of a PDN's domain.   For example, when hospitals share health records, they typically protect patient privacy with the heuristics of HIPAA Safe Harbor~\cite{safeHarbor}.  

We hypothesize that with a combination of common-sense rules and external incentives (e.g., legal remedies for bad actors), we can make a much larger set of data available for sharing.  One of the questions we are pursuing with this work is how far a rule-based PDN security policy will go in a real-world setting.    Going forward, if users wish to add rigorous tuple-level privacy to their SQL workflow, they may integrate differentially private querying into a PDN  using the techniques described in~\cite{mcsherry2009privacy}.


{\sc \sysname} is a substantial departure from existing research on secure databases.  Prior work used homomorphic encryption for the outsourced storage and querying of private data~\cite{Popa2011}.   Our approach keeps data in the hands its originators and uses SMC to conceal the query's computation rather than the data itself.  Researchers have also studied secure computation for outsourced database systems~\cite{aggarwal2005two,chow2009two}.   In contrast, a PDN distributes secure computation among the data providers and removes the need for any trusted intermediaries beyond a lightweight honest broker for coordinating query evaluation.

There are numerous existing approaches to making SMC available to untrained users, including domain-specific programming languages ~\cite{Liu2015,malkhi2004fairplay,rastogi2014wysteria} and extending existing languages~\cite{zahur2015obliv}.   Our approach differs from them in that we do not require the programmer to reason explicitly about how to combine the data of each party.  Instead we use SQL's semantics to translate queries into SMC.   To the best of our knowledge, this is the first system that enables users to take advantage of SMC without reasoning about the security properties of the underlying system.  We decouple the security policy from the querier by providing the honest broker with a schema-based security policy at startup. 


Our main contributions in this work are:
\begin{itemize}
\itemsep1pt \parskip0pt \parsep0pt
	\item A novel generalization of federated databases for querying over data providers that do not trust one another.
	\item A code generator that automatically translates SQL into secure computing primitives.
	\item A heuristics-based optimizer for PDN query execution
	\item An in-depth evaluation of this system on real-world medical data.
\end{itemize}

The remainder of this paper is organized as follows.  In Section~\ref{sec:background} we describe the basics of SMC and introduce a running example for PDNs.  Section~\ref{sec:overview} provides an overview of the {\sc \sysname} system.  Next, we describe our code generator in Section~\ref{sec:code-gen}.  Section~\ref{sec:priv-model} describes our PDN security policy and how  we use it to identify the subtree of a query plan for which SMC is needed.   We then explore our secure query plan optimizations.
  After that, we present  the results our experimental evaluation over real-world  data.  
Finally, we survey the relevant literature and conclude.

\section{Background}
\label{sec:background}

{\sc \sysname} leverages existing SMC primitives for secure query evaluation.
In this section, we briefly describe these building blocks and give an intuition about how they work.  We also introduce a running example from clinical data research.  For clarity, we refer to two mutually distrustful data providers as Alice and Bob.

\subsection{Secure Multiparty Computation}
\label{sec:smc}
SMC systems allow parties to jointly compute functions while keeping each party's inputs secret. In PDNs, database operators act as secure functions that are evaluated over the sensitive data of two or more parties. {\sc \sysname} performs secure computation using garbled circuits, and conceals its access patterns of sensitive tuples with oblivious RAM. We chose these classical techniques because they are heavily optimized in existing work and easily accessible for code generation.  

\noindent {\bf Garbled Circuits}  For secure query evaluation, we generate garbled circuits to compute database operators over the data of multiple parties.  Garbled circuits are a series of logic gates (e.g., AND, XOR) that hide the program traces of their computation.  These circuits use the same distribution of execution time, regardless of their inputs, to make it impossible for a party to deduce the inputs of others.  In our framework, the only information available to the data providers is the number of tuples provided by each party.     Garbled circuits are quite expressive, and can be used to compute any arbitrary function~\cite{Chaum1988}.  This technique has been extended for  three or more parties~\cite{Goldreich1987}. 

Each garbled circuit securely computes a function, such as $a \geq b$ in Yao's Millionaire Problem.  In order to not reveal the inputs of Alice and Bob, we execute the entire circuit in the worst-case performance scenario.  If we were computing $a \geq b$ in plaintext, we'd find the most significant bit where Alice and Bob's inputs differ and use it to determine the circuit output.  A garbled circuit must instead compare all of the bits in order to not disclose the most significant one where the inputs differed.   In {\sc \sysname}, the first data provider, Alice, generates the garbled circuits, and the second one, Bob, evaluates them.  


\noindent {\bf Oblivious RAM}  In addition to covering the compute traces effected by the input of another party, our engine hides the memory access patterns of a secure program.  We use {\it oblivious RAM} (ORAM)~\cite{Goldreich:oram} to store arrays of tuples as they move through the secure query executor.  This data structure 
shuffles the tuple array at each read or write, thereby making all memory accesses indistinguishable from one another.  This prevents attackers from using reads and writes in a secure program to learn about the underlying data.   ORAM enables us to reduce the depth of our garbled circuits by creating a small circuit for each database operator and securely passing the output of one operator to the next through oblivious reads rather than evaluating the query in a single, massive circuit.  \cut{We use the techniques in~\cite{Liu2014} to automatically split up the circuits. This oblivious data structure makes our system practical, but it comes at a high cost.  Each read or write with ORAM takes polylogarithmic time, O($(log~ n)^3$), where $n$ is the number of elements in the array.}

\subsection{HealthLNK Running Example}
\label{sec:running-example}

Throughout the text, we use a running example of a group of  hospitals that wish to mine the collective data of their electronic health record systems for research while keeping individual tuples private.  We first examine the architecture of this system and then describe a set of representative queries with which we explore {\sc \sysname}.

A clinical data research network or CDRN is a consortium of healthcare sites that agree to share their data for research.  CDRN data providers may be mutually distrustful parties.   We examine this work in the context of HealthLNK~\cite{healthlnkirb},  a CDRN prototype for Chicago-area healthcare sites.   This repository contains records  from seven Chicago-area healthcare institutions, each with their own member hospitals, from 2006 to 2012, totaling about 6 million records.  The data set is selected from a diverse set of hospitals, including academic medical centers, large county hospitals, and local community health centers.  

HealthLNK is the forerunner for the Chicago Area Patient-Centered Outcomes Research Network (CAPriCORN), which is itself part of a national network in the US, the Patient-Centered Outcome Research Network (PCORnet).   The CAPriCORN consortium includes 481 data sources, each of which has protected health information (PHI) such as gender, timestamps, and diagnoses~\cite{kho2014capricorn}.  
CAPriCORN and HealthLNK share the majority of their stakeholders and this group designed these systems to meet the needs of clinical researchers, especially ones exploring personalized medicine.


 In the absence of a secure query evaluation framework like {\sc \sysname}, CDRN members resort to computing distributed queries entirely within the honest broker.   Because each data provider is in charge of their own HIPAA compliance, they will often only volunteer their least sensitive database attributes for querying.  Although each site will not disclose its PHI, they are willing to compute queries over sensitive data when multiple records are accessed together.  

In addition to limiting the queryable attributes in a CDRN, this hub-and-spoke architecture will not scale to hundreds of data providers.  This is an issue we are already seeing in the field.  On the other hand, this setting makes it is possible for individual tuples to ``hide in the crowd'' of the data from many participating healthcare sites.   This makes CDRNs a prime use case for {\sc \sysname}.   We are investigating deploying a prototype of {\sc \sysname} on CAPriCORN.

\subsubsection{\bf Query Workload} 
We now explore the workings of {\sc \sysname}  with  three representative queries based on clinical data research protocols~\cite{hernandez2015adaptable, cdiff2015irb}.  In addition, we evaluate the system with this workload on de-identified medical records from the HealthLNK data repository.  
%

\noindent  {\bf Comorbidity} {\it Clostridium difficile}, or {\it c.\@ diff}, is an infection that is often antibiotic-resistant.   Our first query finds the most common types of ailments that arise for {\it c.\@ diff}  sufferers:
\begin{verbatim}
SELECT diag, COUNT(*) cnt
FROM diagnoses 
WHERE patient_id IN cdiff_cohort
GROUP BY diag 
ORDER BY cnt
LIMIT 10;
\end{verbatim}
The query selects the diagnoses of individuals in the {\it c.\@ diff} cohort and counts the diagnoses for each condition, returning the most common ones to the user.  With {\it comorbidity}, we explore practical techniques for minimizing the use of SMC in distributed query evaluation.

\noindent {\bf  Recurrent C.\@ Diff} {\it C.\@ diff} sufferers have a high rate of re-infection after treatment.  When patients are treated at multiple hospitals, recurrent {\it c.\@ diff}  frequently goes undetected.  This is exactly the type of problem that {\sc \sysname} is designed to solve.  This query identifies a cohort of recurrent {\it c.\@ diff} patients whose second infection occurs between 15 and 56 days after an initial diagnosis:
\begin{verbatim}
WITH rcd AS (
    SELECT pid, time, row_no() OVER 
        (PARTITION BY pid ORDER BY time)
    FROM diagnosis
    WHERE diag=cdiff)

SELECT DISTINCT pid
FROM rcd r1 JOIN rcd r2 ON r1.pid = r2.pid
WHERE r2.time - r1.time >= 15 DAYS
     AND r2.time - r1.time <= 56 DAYS 
     AND r2.row_no = r1.row_no + 1;
\end{verbatim}
We first select for {\it c.\@ diff}, and use a window aggregate to number the diagnoses of each patient in chronological order.
We then compare the $i$th diagnoses to the $(i+1)$th one using a self-join to find recurring infections in the prescribed date range. Lastly, the query eliminates duplicate patient IDs.  We use this query to examine how fine-grained data partitioning (by patient ID) improves {\sc \sysname}'s performance.

\noindent {\bf Aspirin Count} In our final query, we identify the number of  heart disease  sufferers who were prescribed Aspirin.  Here, researchers are investigating the effectiveness of Aspirin in preventing repeated heart attacks.   We calculate the {\it Aspirin Count} of heart disease patients as: 
\cut{If patients receive a heart disease diagnosis and an Aspirin prescription from different care providers, this prevents researchers' attempts to link an attack with its treatment. } 

\begin{verbatim}
SELECT COUNT(DISTINCT pid)
FROM diagnosis d JOIN medication m ON d.pid = m.pid
WHERE d.diag = hd AND m.med = aspirin 
    AND d.time <= m.time;
\end{verbatim}

The query first filters the diagnosis table for heart disease patients and the medications for Aspirin.  It then joins these tables to identify patients who were prescribed Aspirin during or after a heart disease diagnosis.  Lastly, we count the distinct patient IDs that meet this criterion. This query tests the {\sc \sysname} optimizer's ability to create high-performance query plans for complex sequences of operators. 



\section{System Overview and Roadmap}
\label{sec:overview}

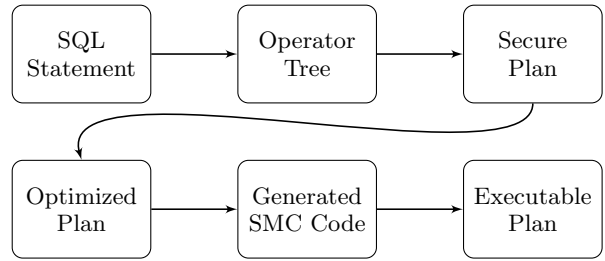
\begin{figure}
\begin{tikzpicture}[node distance=3cm]
\tikzstyle{block} = [rectangle, draw, 
    text width=5em, text centered, rounded corners, minimum height=4em]
\tikzstyle{line} = [draw, -latex',semithick]

\node (sql) [block] {SQL Statement};
\node(optree) [block, right of=sql]{Operator Tree};
\node(secmodel) [block, right of=optree] {Secure Plan \cut{\small  Section~\ref{sec:priv-model}}};

\node(opt) [block, below=0.75cm of sql] {Optimized Plan \cut{\small Section~\ref{sec:optimization}}};
\node(codegen) [block, right of=opt] {Generated SMC Code \cut{\small Section~\ref{sec:code-gen}}};
\node(execplan) [block, right of=codegen] {Executable Plan};

\path [line] (sql) -- (optree);
\path [line] (optree) -- (secmodel);
\path [line] (opt) -- (codegen);
\path [line] (codegen) -- (execplan);

\path[line] (secmodel.south) .. controls +(down:11mm) and (left:4mm) ..(opt.north);

\end{tikzpicture}
\caption{{\sc \sysname} query compilation and execution}
\label{fig:overview}
\vspace{-5mm}
\end{figure}

In this section, we walk through the steps {\sc \sysname} takes to translate a SQL statement into a secure query execution plan, as detailed in Figure~\ref{fig:overview}.  
The honest broker starts with a SQL statement provided by the user.  The statement is written against the PDN's shared schema.  The honest broker parses the statement into a directed acyclic graph (DAG) using well-known techniques~\cite{ramakrishnan2000database}.  This tree of database operators, such as joins and aggregates, provides the steps needed to compute a given query.  The honest broker examines this tree, confirming that it is runnable within the PDN's security policy.  

Next, we generate a {\it secure plan} that identifies the minimal subtree in the query's DAG that must run obliviously to uphold the PDN's security policy.    We describe this process in Section~\ref{sec:sec-model}.  The planner traverses the tree bottom-up, modeling the flow of sensitive attributes through its nodes.  
 \cut{ resolving all of the attributes in its leafs to their security policy and } 

Next, as described in Section~\ref{sec:optimization}, {\sc \sysname} optimizes the secure query tree using heuristics that partition database operators into small, scalable units of secure computation.  We also propose methods for reducing the secure computation performed within an operator.   

Armed with a tree of optimized operator specifications, the planner generates SMC code for execution on the data providers using the techniques in Section~\ref{sec:code-gen}.   For each relational algebra operator, the code generator looks up a template for it and populates this outline with query-specific information including the width of its tuples in bits and filter predicates. 

When the code generator completes we have an executable secure plan.  The honest broker distributes the compiled secure code to the data providers, along with plaintext source queries for generating inputs for SMC.  The data providers run their source SQL queries within their databases with standard methods and coordinate with one another to generate and evaluate garbled circuits for the secure operators using the specifications provided by the SMC code.

End-to-end, {\sc \sysname}  implements a wide range of SQL operators.  It supports selection, projection, aggregation (including {\tt DISTINCT}), limit, and some window aggregates.  For joins, we handle equi-joins, theta joins, and cross products.  At this time we do not support outer joins or set operations.  We detail how the latter's semantics would be implemented in Section~\ref{sec:sec-model}.


\vspace{-3mm}
\section{Secure Query Executor}
\label{sec:code-gen}

We now examine the process whereby {\sc smcql} translates physical operators in a PDN plan into executable, secure code.  We first prepare the plan by adding steps that combine tuples from multiple parties.  After that, we generate secure code for each relational operator in the operator tree.   



This secure code uses garbled circuits to jointly compute an operator over the participating parties, and stores the result in ORAM.  The engine uses this secure data structure to pass intermediate results between nodes in the operator tree.   The query executor carries out this secure computation\textendash with Alice generating the circuits and Bob evaluating them\textendash for every subsequent operator in the tree until it reaches the root node.  Each party ships their garbled result from the root node to the honest broker for decoding the query's output tuples.

To create garbled circuits and ORAM for PDN query evaluation, we use a domain-specific programming language, ObliVM~\cite{Liu2015}.  This language has a C-style syntax.  Among other features, ObliVM offers callable functions, loops, and if-then statements.  A programmer declares and accesses ORAM in ObliVM using the C language's bracket notation.     This framework compiles its code in two steps.  First, it translates the code into a set of logic gates and ORAM accesses.   Then, at execution time, it generates the garbled circuits on the fly to prevent replay attacks.  This language is backend-agnostic, so if one were to research new garbled circuit protocols, they would be able to seamlessly compile them into  the code generated by {\sc \sysname}.

Each secure operator starts with a template, or a parameterized C-style program for the operator's execution logic.   Templates have variables for filter predicates, input tuple widths, and for projecting their output as needed.  The system has a library of operator-specific templates, including ones for the optimizations introduced in Section~\ref{sec:optimization}.   

We show an example template for joins in Figure~\ref{fig:join-template}.  The template's parameters are denoted with a ``\$''.    The system populates the template's variables using parameters from its relational query plan.   This join function takes in two arrays of tuples, one for each input.  Each array contains tuples from both parties.    The variables \$lSize and \$rSize denote the number of bits per tuple in each input relation.  The tuples per input table are stored in $n$ and $m$, and the SMC executor infers these values at runtime.

After populating an operator's template, we compile it into a low-level representation with logic gates and ORAM accesses.  This is the code that each data provider executes.  These low-level directives automatically generate all of the garbled circuits  (and their coded inputs) at runtime to protect the secure  query evaluation from replay attacks.




\begin{figure}
{\scriptsize
\begin{verbatim}

int$dSize[m*n] join(int$lSize[m] lhs, int$rSize[n] rhs) {

	   int$dSize[m*n] dst;
	   int dstIdx = 0;
	
   	for(int i = 0; i < m; i=i+1)  {
   	   int$lSize l = lhs[i];
   	   for(int j = 0; j < n; j=j+1) {
             int$rSize r = rhs[j];
	             if($filter(l, r) == 1) {
	                  dst[dstIdx] = $project;
		                  dstIdx = dstIdx + 1;
	             }
	        }
	   }
	   return dst;
}\end{verbatim}}
\vspace{-5mm}

\caption{Template for secure join}
\label{fig:join-template}
\vspace{-5mm}
\end{figure}

\subsection{SMC Performance Costs}
\label{sec:handcoded}
All SMC techniques incur substantial overhead in comparison to their plaintext counterparts.   To get an intuition for the performance costs and optimization opportunities associated with secure query evaluation, we hand-coded the three queries in Section~\ref{sec:running-example} as oblivious programs.  We ran these carefully optimized, fully-SMC implementations on a randomly selected subset of tuples that matched the query's initial selection criteria.  The details of our experimental configuration are in Section~\ref{sec:experiment-config}. \cut{For example,  the {\it comorbidity} query operated on co-occurring diagnoses from {\it c.\@ diff} sufferers. } To bound the duration of our experiments, our input samples had 50 tuples per table.

Our results are shown in Table~\ref{tbl:handcoded}. We see that a purely secure query evaluation is on the order of 4\textendash 5 orders of magnitude slower than its plaintext execution. This is unacceptably slow. We see that for complex plans with cascades of operators, like {\it aspirin count}, the runtime explodes.  In order to keep their computation oblivious, each step in the query has an output cardinality equal to its maximum possible size.  For example, the output of a join is the size of the cross product of its inputs.  Thus {\it aspirin count}'s aggregate after the join must process all of these tuples\textendash despite many of them being nulls\textendash to avoid revealing information about the contents of the join inputs.

Instead of na\"\i vely using SMC on the entire query plan, we need a more nuanced approach. 
The {\sc \sysname} optimizer in Section~\ref{sec:optimization} uses SQL semantics to minimize the work performed in SMC and it breaks this secure computation into small, scalable partitions for speedy evaluation.  These optimizations are agnostic to the SMC primitives used in distributed query evaluation.  Selecting the highest-performance SMC protocol for a given query is a promising direction for future work. Before we can discuss our optimizations, we first introduce a security model with which we identify query evaluation over sensitive data. This model allows us to formally describe our optimizations and verify that the system satisfies  the security requirements of its stakeholders.

\begin{table}
\centering
\caption{Slowdown of HealthLNK queries run with SMC.}
\begin{tabular}{|l|c|c|c|}
\hline
Test & Plaintext & Secure & Slowdown \\
\hline
{\it Comorbidity} & 158 & 253,894 & 1,609X \\
{\it Recurrent C.\@ Diff} & 165 & 159,145 & 967X \\
{\it Aspirin Count} & 193 & 8,195,317 & 43,337X \\
\hline
\end{tabular}
\label{tbl:handcoded}
\vspace{-5mm}
\end{table}

\section{Security Model}
\label{sec:priv-model}
We now formulate the challenge of creating oblivious query execution plans.  First, we describe {\sc \sysname}'s user-facing security policy for PDN queries.  This attribute-level model specifies who may access the PDN's data and under what conditions.   After that, we explore {\sc \sysname}'s security type system, with which the planner identifies the minimal set of operators in a query plan that require oblivious evaluation to protect a PDN's sensitive data.  

\subsection{Security Policy} 
\label{sec:schema}

{\sc \sysname} offers a simple, yet powerful security model to protect PDN data from unauthorized access.  Recall that a PDN begins with a common set of table specifications and that they define the level of protection needed for their data one column at a time. This approach mirrors how PDN  stakeholders often reason about the sensitivity of their data.  

Our model offers three levels of  data access: public, protected, and private.  By working with stakeholders, a DBA creates an annotated schema that specifies the PDN's security policy.   This policy  also enables the {\sc \sysname} optimizer to create efficient, secure plans.

{\em Public} attributes are readable by all parties, including the honest broker, data providers, and end users.   These columns  have the lowest sensitivity, and often have a low probability of being independently replicable.  In HealthLNK, anonymized patient IDs, lab results, and blood pressure readings are public attributes.

{\it Protected} data is visible in their originating site and conditionally available to the end user and honest broker.   {\sc \sysname} uses  $k$-anonymity to control access to protected attributes.  A selection is $k$-anonymous iff each of its tuples is indistinguishable in its protected attributes from at least $k-1$ records.  This policy is one of many possibilities for controlling access to protected data.   Any distributed query evaluation  over these attributes is done securely.  In our running example, protected attributes include diagnosis codes, age, and gender.  \cut{ It is possible to configure a PDN with multiple classes of protected attributes, each with a different access control policy.}

{\em Private} attributes are the most sensitive ones in a PDN, and they are not disclosed to anyone outside of the initial data provider.  Computation over these attributes must be carried out obliviously. Private attributes may not appear in any results returned to the user.  Timestamps and zip codes are examples of private attributes in HealthLNK.

\cut{When multiple attributes are accessed in a query together, {\sc \sysname} operates at the security level of the most sensitive attribute.  Before the framework starts executing a query, it validates the operation against the security policy provided by the annotated schema.  The framework uses this policy to plan PDN queries that will run efficiently by minimizing the computation it performs within SMC.}

This access control policy governs when and how a PDN uses secure computation. \cut{by analyzing the underlying data accessed by a query. }  In addition, the PDN is configured with a query admission policy.
This policy may disallow certain patterns of querying, such as repeated, but slightly modified ones designed to unmask individuals in a database.  It may also enable data providers to hide in the crowd with requirements such as ``at least $k$ data providers must contribute tuples to a secure computation''.   A PDN may automatically reject queries that do not meet its policy,\cut{as specified by the stakeholders of the network,} using a system such as DataLawyer~\cite{upadhyaya2015automatic}.  Another approach is to audit query trails to determine if a sequence of queries is threatening to unmask sensitive views of the data~\cite{motwani2008auditing}. \cut{We leave the design of PDN query admission to future work.  }

We now address the planning and optimization of PDN queries.  Since query processing takes place within the data providers, we treat protected  attributes as if they are  private.  




\subsection{Secure Information Flow}
\label{sec:sec-model}

Our first tactic for optimizing a PDN query is to minimize the number of  operators it runs securely.    We use a security type system~\cite{Kerschbaum2013,sabelfeld2003language,schneider2001language, volpano1996sound} to analyze the flow of secure attributes through an operator tree.   The planner traverses the tree bottom-up, recording the operators that will be executed obliviously to fulfill the requirements of the PDN's security policy.   In each operator, we examine the provenance of its output columns and determine the protection level needed for each  one by  taking the maximum security policy of its source attributes.

\cut{When crafting a secure query plan, our first goal is to guarantee that its execution will not reveal information to any party that exceeds their permissions.  Thus data providers cannot learn anything about the contents of the private and protected attributes of their peers.  Likewise, the honest broker can learn nothing about private data except that which they can surmise from the output of a query.  

We identify when secure computing is needed in a query plan by tracing the flow of private attributes through operators in the query tree. }

The security type system begins with a grammar with which the planner interprets the query's operator tree.  Grammar~\ref{gram:sql-syntax}  shows the syntax with which the type system will analyze a query tree to determine whether each of its database operators needs oblivious evaluation.  This syntax closely follows that of relational algebra.  


\grammarindent=80pt

\begin{Grammar}
 \begin{grammar}
<phrases> $\rho$ ::=  e | E | Op

<expressions> e ::= a | n | e + e' | $e \leq e' $ | $e \land e'$ | ...

<sets> E ::= \{$e_1, e_2, \ldots e_n$\}

<operators> Op ::= Op'(Op(.)) \\ | Op'(Op(.), Op(.)) | {\tt scan(}E{\tt )} \\| $\sigma_{e}(E)$  | $\pi_{E'}(E)$ | $E \bowtie_e E'$ \\ | {\tt  agg(}E{\tt )} |  {\tt  limit(}E{\tt )} | {\tt sort(}E{\tt )} \\ |    $E \cup E'$ | $E \cap E'$ | $E \setminus E'$
\end{grammar}  

\caption{Grammar for SQL query plans.}
\label{gram:sql-syntax}
\end{Grammar}

All objects in a query plan are phrases, represented by $\rho$.  A phrase may be an expression ($e$), a set of expressions ($E$), or a relational operator ($Op$).  Expressions may describe attribute references ($a$), string and integer literals ($n$), arithmetic operators, comparisons or logical connectives.    

We reason about a query plan as a set of relational operators.   Operators are arranged in a tree and each operator has up to two children.  An operator takes in a set of expressions, E, and produces a new set, $E'$, as output.  The grammar offers table scans, filters ($\sigma_{e}(E)$), projections ($\pi_{E'}(E)$), joins ($\bowtie$), aggregates, sorts, limit, and set operations.


\grammarindent=80pt

\begin{Grammar}
 \begin{grammar}

<security types> $\tau$ ::= $s \in \{low, high\}$

<phrase types> $\rho$ ::= $\tau $ |   $\tau \: set$ |  $\tau \: exec$
\end{grammar}  

\caption{Grammar for secure information flow analysis}
\label{fig:sec-syntax}
\end{Grammar}

The security type system assigns a {\it label} to each phrase in a query plan.  To model the flow of secure attributes through the query tree, we label each phrase as {\it low} or {\it high}.   If a  phrase is {\it low}, it does not require oblivious computing and we run it within the source databases like a conventional federated query.  A {\it high}  phrase requires oblivious evaluation.   We use low to denote computing over public attributes or in a setting where secure evaluation is not needed.   Private attributes are handled with {\it high} operators and we call the set of {\it high} attributes in a PDN schema $h$.  In addition to judging each phrase as {\it low} or {\it high}, the system records the type of each phrase to show whether it is referring to an expression (the default), set of expressions, or an operator execution.  This syntax is shown in Grammar~\ref{fig:sec-syntax}.


For each typing rule, we say
$$ \frac{assumptions}{type\; judgement}$$

\noindent to denote the conditions under which we assign a security label to a given phrase.  Each type judgement rule is of the form $\gamma \vdash \rho : \tau \; type$.  This rule says that in type system $\gamma$, we judge phrase $\rho$ as security type $\tau \in \{high,\, low\}$, with a phrase type of expression, set, or execution.

Figure~\ref{fig:exp-type-rules} shows the rules for labeling expressions and sets thereof.    In rule {\sc e-base} we say that any phrase may be evaluated as {\it high}.   An expression may be labelled as {\it low} iff none of the attributes referenced in it are {\it high}.  In {\sc e-set} we assign a type to a set of expressions, $E$,  by resolving its elements to a single label, $\tau$.  If $E$ contains a mix of {\it low} and {\it high} expressions, the system uses type coercion to assign a {\it high} label to the set.

\begin{figure}
\begin{tabular}{ll}
 {\sc e-base} & $\gamma \vdash e : high$  \\[0.5cm]

{\sc e-low} & $\inferrule{h \not\in attrs(e)}{\gamma\, \vdash\, e \,: \,low}$\\[0.5cm]

 {\sc e-set} & \inferrule{E = \{e_1,\ldots,e_n \}
 \\\\
\forall i \in \{1\ldots n\} :   (\gamma \vdash e_i : \tau)}{\gamma \vdash E : \tau \; set} 

\end{tabular}
\caption{Rules for  information flow in relational data.}  
\vspace{-5mm}
\label{fig:exp-type-rules}
\end{figure}

The type system rules for labeling relational algebra operators in a query plan are shown in Figure~\ref{fig:sql-type-rules}.  We classify each relational algebra operator into one of two categories: tuple-at-a-time or multi-tuple evaluation.   Most tuple-at-a-time operators, or ones that emit output by considering each tuple discretely, have no need for secure evaluation unless they consume data from one or more secure children.  Operators of this kind are scan, SQL project (for altering the attributes in intermediate results), limit, and filters with {\it low} predicates.  As shown in {\sc r-filter}, a filter needs to be evaluated obliviously iff its predicate changes the operator's output cardinality based on private attributes.   Scans are unconditionally public because they are executed locally and  their output cardinality is not altered by any attributes.

\begin{figure}[!ht]
\begin{tabular}{ll}
  {\sc r-aggregate} & \inferrule{\gamma \vdash E : \tau\, set}
{\gamma \vdash agg(E) : \tau \, exec}\\[0.5cm] 
 {\sc r-distinct} &   \inferrule{\gamma \vdash E : \tau\, set}
{\gamma \vdash distinct(E) : \tau \, exec}\\[0.5cm] 

  {\sc r-filter} & \inferrule{\gamma \vdash exp : \tau }{\gamma \vdash \sigma_{exp}(E) :  \tau \, exec}\\[0.5cm]

 {\sc r-join}  & \inferrule{\gamma \vdash E : \tau\, set
\\\\
\gamma \vdash E' : \tau\, set}
{\gamma \vdash E \bowtie E' : \tau \, exec} \\[0.5cm]

  {\sc r-scan} & $\gamma \vdash scan(E) : low\, exec$\\[0.5cm]

{\sc r-setop}  &  \inferrule{\gamma \vdash E : \tau\, set
\\\\
\gamma \vdash E' : \tau\, set}
{\gamma \vdash E \, op\, E' : \tau \, exec}  \\[0.5cm]
{\sc r-sort}  & \inferrule{\gamma \vdash E : \tau\, set}
{\gamma \vdash sort(E) : \tau \, exec}\\[0.5cm] 


{\sc r-nest} &\inferrule{\gamma \vdash Op(.) : \tau\, exec 
\\\\
\gamma \vdash Op'(.) : \tau\, exec
}{ \gamma \vdash Op'(Op(.)) : \tau \, exec}\\[0.5cm]

 {\sc r-nest-bin} &\inferrule{\gamma \vdash Op(.) : \tau\, exec 
\\\\
\gamma \vdash Op'(.) : \tau\, exec
\\\\
\gamma \vdash Op''(.) : \tau\, exec
}{ \gamma \vdash Op''(Op(.), Op'()) : \tau \, exec}

\end{tabular}
\caption{Typing system rules for  information flow in relational algebra query tree.}
\label{fig:sql-type-rules}
\end{figure}

Multi-tuple operators\textendash including sorts, joins, and set operations\textendash combine data from multiple source engines, and they execute at the level of their most sensitive inputs.  The type judgements of this kind are {\sc r-aggregate}, {\sc r-distinct}, {\sc r-join}, {\sc r-setop}, and {\sc r-sort}.   This, in conjunction with the nesting rules, ensure that these operators leak no information about private data.

All additional operators are covered by {\sc r-nest}; it states that a new operator $Op'$ executes at a security level greater than or equal to that of its child.  In {\sc r-nest-bin}, we generalize this to operators with two inputs, i.e., joins and set operations.  If a binary operator has at least one {\it high} child, then the operator is judged as {\it high}.  The nesting judgements ensure that no subsequent operators leak information about a prior secure computation.

\noindent{\bf Security Proof} We prove the security of our information flow type system by induction.  The type system traverses a query tree bottom-up.

\noindent {\sc Base Case}: The leafs of the query tree are all table scans.  By {\sc r-scan} every branch of the tree begins as {\it low}.  Table scans are unconditionally oblivious\textendash the contents of the tuples do not alter the data flow of this operator.

\noindent{\sc Step 1}: There are two cases for the parent operator of a scan.  The operator may process data one tuple at a time.  Recall that projects, limits, and {\it low} filters are in this category.  By {\sc r-filter}, the filter operator ceases to be oblivious when its predicate references non-public attributes.  Otherwise, the filter executes at the same security level as its child by {\sc r-nest}.    Likewise, SQL project and limit operators are deterministic in the cardinality of their outputs, so they reveal no information that is not already visible from running them at the security of their source operator.   

The remaining operators\textendash aggregation, distinct, join, set operations, and sort\textendash compute  over multiple tuples.  Thus, they may execute over the data of mutually distrustful parties.   Each of these operators executes at the security level of the most sensitive attribute in their input expressions.  \cut{They do this in two steps.}  By {\sc e-set}, the type system determines the security label of the expressions in an operator's fields.  The resolver labels the operator with the set's security level, $\tau$, using the operator-specific rules in Figure~\ref{fig:sql-type-rules}.

\noindent{\sc Step $n$}: Each subsequent step executes at a security level greater than or equal to that of its children.  By {\sc r-nest} and {\sc r-nest-bin}, an operator is typed with the label of its source nodes.  Therefore, even if an operator only references public attributes, such as  {\tt DISTINCT patient\_id} in {\it recurrent c.\@ diff}, the plan reveals no information about the output of secure computation performed by its source operators.

\section{Secure Query Optimization}
\label{sec:optimization}

Armed with our security model, we can now examine optimizations to improve the performance of {\sc \sysname} for secure querying. The optimizer begins with a secure query plan, wherein each operator is labelled {\it high} or {\it low} using the type checker in Section~\ref{sec:sec-model}.  It then transforms this logical plan into a low-level physical one.  Using heuristics, {\sc \sysname} reduces a secure plan's use of secure computation and partitions query evaluation into small, scalable units.  The optimizations described below are generalizations of distributed query optimization~\cite{chaudhuri1998overview}. Since these adjustments change the program structure of a secure plan, we extend the security proof in Section~\ref{sec:sec-model} for them.

\subsection{Scalable Physical Plans}
\label{sec:slicing}

The optimizer identifies opportunities to {\it slice} its secure query evaluation by partitioning the input data into smaller, more manageable units of computation.  This fine-grained partitioning enables us to reduce our secure code complexity, thereby  speeding up the secure computation of eligible queries.  Slicing also makes it possible to parallelize secure query evaluation.  The optimizer assigns each operator in a secure plan  to one of three execution modes:
\begin{itemize}
\itemsep1pt \parskip0pt \parsep0pt
\item Plain:  Operators are of type {\it low} and they evaluate in the source database.
\item  Sliced: {\it High} operators run securely  over tuples that are horizontally partitioned by a public expression.   
\item Secure:  {\it High} operators executed using a single SMC program run on the inputs of all data providers.
\end{itemize}

\cut{The execution modes are ordered from lowest to highest in their security requirements. As tuples move up the query tree, their execution mode monotonically increases.  Thus} All paths in the tree start with plaintext scans over a database table.  When the executor encounters a {\it high} operator, the engine  switches to sliced or secure execution mode.  If an operator is in sliced mode, then its ancestors must run in sliced or secure mode so that the query's computation remains oblivious.

 Each sliced operator that receives plaintext data bins its input by a {\it slice key}, or an expression on public attributes upon which the engine divides up a secure operator's computation.   Each distinct value associated with a slice key, or {\it slice partition},  is computed independently.  For each relational algebra operator in Grammar~\ref{gram:sql-syntax}, we identified if and how it is sliceable.   

Joins with public predicates are runnable one slice partition at a time.  Likewise, it is possible to slice sorts by all or part of their sort key.   Sliced aggregates compute one group-by bin at a time and {\tt DISTINCT} operators break up their computation by the attributes they reference.  Since filters and projections work one tuple at a time, they are agnostic to slicing and assume the slice key of their parent or child operator, whichever enables more sliced evaluation.  Set operations with any slice key will produce the correct outputs because a non-empty key guarantees that any tuples needing comparison will be grouped together.

Slicing is a powerful optimization because it is composable.  The optimizer identifies sequences of secure operators that partition their computation on the same slice key.  For example, in the  {\it recurrent c.\@ diff} query every operator after its initial table scan is computed securely and all oblivious operators are sliced by patient ID.

\begin{algorithm}
\begin{algorithmic}
\small 
\STATE \textbf{Function} \textbf{planExecution}
\STATE \textbf{Input:} DatabaseOperator $o$ 
\STATE \textbf{Output:} ExecutionMode $e \in \{Plain, Sliced, Secure\}$ 
\STATE $e$ = Plain;
\IF{o.label = {\it low}} 
\STATE return e; 
\ENDIF

\FOR{$c \in$ o.children()} 
	  \STATE childMode = planExecution(c);
	  \IF{childMode == Secure}
		\STATE $e$ = Secure;
	   \ELSIF{childMode == Sliced}
		\IF{o.sharesSliceKey(c) \AND $e \neq$ Secure}
			\STATE $e$ = Sliced;
		\ELSE
			\STATE $e$ = Secure;
		\ENDIF
	\ENDIF
\ENDFOR

\STATE // if  o.label = {\it high} and all children computed in plaintext
\IF{$e$ == Plain  \AND o.sliceKey $\neq \emptyset$}
	\STATE $e$ = Sliced;
\ELSE 
	\STATE $e$ = Secure;
\ENDIF
\RETURN $e$
\end{algorithmic}

\caption{Method for assigning execution mode to database operators in a secure  query plan.}
\label{a:inferExec}
\end{algorithm}

The optimizer identifies sequences of  sliced computation by traversing a query tree bottom-up using Algorithm~\ref{a:inferExec}.  All {\it low} operators are run in plaintext. 
If a {\it high} operator has only plain children and a nonempty slice key, then it is assigned to slice mode.   
The optimizer then checks to see if the operator shares a slice key with its parent. Two operators are sliced alike if their slice key is equal, or for joins, at least one side of an equality predicate appears in the slice key of each of its descendants.   If a {\it high} operator does not qualify for slicing, we switch to secure mode.

\noindent{\sc Security Proof} Each sliced query evaluation is equivalent to running a secure operator without slicing where we insert a filter over public attributes for each distinct slice partition.   The optimizer identifies partitions of computation by running {\tt SELECT DISTINCT} {\it $<$slice key$>$} {\tt FROM ... } on each data provider.  By {\sc r-distinct} this query is public.  After that, we evaluate each slice partition securely using the previous plan with an added filter for each slice partition.  By {\sc r-filter} this filter is public, thus we retain the security properties of the previous plan.

\subsection{Minimizing Secure Computation }
\label{sec:odf}


We now introduce optimizations that reduce a query's use of secure computation.  The first delays its entry into SMC  by partially computing an operator in a lower execution mode.  The second reduces the data that the engine evaluates securely.  

\noindent{\bf Split Operators}   A {\it splittable} operator may partition its execution into discrete phases, such as local plaintext followed by distributed secure computation.   For example, if we are computing a {\tt COUNT(*)}, the query executor may have each data provider calculate a local partial count and use SMC to add them up.  Aggregates and  filters are splittable operators.


Each splittable operator has a {\it low} phase and a {\it high} phase, and the low phase is executed first.  For aggregates, the {\it low} phase partially computes the aggregate over the input tuples and the {\it high} one combines the partial aggregates.   
Filters with conjunctive predicates are splittable into clauses that reference sensitive attributes and ones that do not.  All other operators are not trivially splittable.

\noindent {\sc Security Proof} For an aggregate with no group-by its output is exactly one tuple, and it is trivially oblivious.   To handle aggregates with a group-by on sensitive attributes, we take advantage of the PDN's architecture to reduce our overhead.  Recall that the parties do not reveal protected or private attributes to one another and that none of the data providers have access to the output of a secure query evaluation.  If an operator computes  over $|A|$  partial aggregates from Alice and $|B|$ from Bob, the output cardinality of its {\it high} phase is unconditionally $|A| + |B|$\textendash with null-padding as needed.    This fixed output length precludes each data provider from learning the group-by values of a partial aggregate that is not their own.   
  \cut{If the aggregate has a group-by, the framework reveals only the number of partial aggregates produced by each provider but not the group-by values associated with them.  Because the output of a secure group-by has a length equal to the sum of the size of its two inputs,  neither of the parties can deduce anything about the distribution of group-by values.  If the honest broker observes the number of tuples provided by each party, they can only deduce the size of the intersection among the parties from the final output.   They learn nothing about the contents of this intersection.   } 
A split filter is analogous to creating two separate selections, one for {\it low} predicates followed by another for {\it high} ones.  We perform a type judgement on each one with {\sc r-filter} and assign its execution mode accordingly.

\cut{\noindent{\bf Precomputed Filters} If we have a filter  that compares expressions with private attributes to literals\cut{, such as {\tt medication LIKE '\%aspirin\%'}}, we need to compute this operator obliviously to protect the data.  On the other hand, evaluating this logic within SMC increases our code complexity thereby slowing down our performance.  To address this, we add a boolean attribute to the initial query selection that represents the outcome of the filter.  We use this stand-in to securely filter the data. For example, we would rewrite:

{\small
\noindent {\tt SELECT patient\_id, date FROM medications\\ WHERE medication LIKE '\%aspirin\%'} 

to

\noindent {\tt SELECT patient\_id, date,medication LIKE '\%aspirin\%'\\ FROM medications}.  
}

\noindent The subsequent secure filter is reduced to a simple check on this new boolean attribute.

\noindent{\sc security proof} Precomputing the filter does not alter the cardinality of the results of the initial data selection.  Since the number of tuples used in secure computation is the same as before the rewrite, this change does not affect the obliviousness of our query plan.
}

\noindent{\bf Secure Semi-Join}  We can further reduce our reliance on secure computation by performing it only on  slice partitions that are present in greater than one data provider.  For example, in {\it recurrent c.\@ diff} if a given patient ID is present in just one hospital, we evaluate the medical records of this individual locally in plaintext, sending the output of this computation to the honest broker over an encrypted channel.    Hence, the query execution plan has two tracks for sliced operators: a secure one for partitions that appear in multiple data providers and plaintext one for all others.

\noindent{\sc security proof} \cut{Semi-join reveals no additional information about the data because the slice keys are comprised of only public attributes.}  To determine the distributed slice partitions, the honest broker collects the distinct slice partitions as in the proof in Section~\ref{sec:slicing}.    It then takes the intersection of their sets.  By {\sc r-setop} the computation of an intersection of two public inputs is public.  By using only public values in our slice keys, this optimization reveals no additional information in comparison to its predecessors.

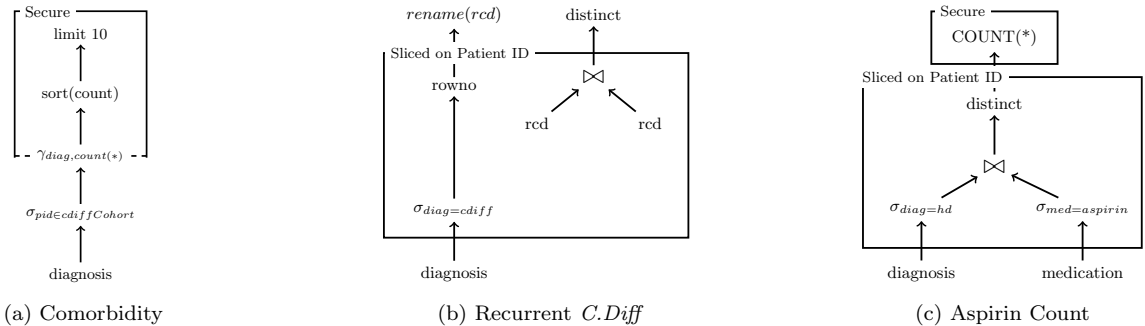
\begin{figure*}[t!]
    \centering
    \begin{subfigure}[t]{0.31\textwidth}
        \centering
	\resizebox {!} {1.5in} {
        	   \begin{tikzpicture}
\node[text centered] (dscan) {diagnosis};
\node[above =  0.75 of dscan, text centered] (filter) {$\sigma_{pid \in cdiffCohort}$};
\node[above =  0.75 of filter, text centered] (agg) {$\gamma_{diag, count(*)}$};
\node[above =  0.75 of agg, text centered] (sort) {sort(count)};
\node[above =  0.75 of sort, text centered] (limit) {limit 10};

\draw[->, line width= 1] (dscan)  --  (filter);
\draw[->, line width= 1] (filter)  --  (agg);
\draw[->, line width= 1] (agg)  --  (sort);
\draw[->, line width= 1] (sort)  --  (limit);

\draw[line width= 1, dashed] (agg.east) + (0.35, 0)  --  (agg.east);
\draw[line width= 1, dashed] (agg.west) + (-0.35, 0)--  (agg.west);

\draw[line width= 1] (-1.35,2.40) --  (-1.35,5.40) -- node [near start,fill=white] {Secure}  (1.35, 5.40) -- (1.35, 2.40);


\end{tikzpicture}
        }
        \caption{Comorbidity}
    \end{subfigure}%
    ~ 
    \begin{subfigure}[t]{0.35\textwidth}
        \centering
	\resizebox {!} {1.5in} {
        	   \begin{tikzpicture}
\node[text centered] (dscan) {diagnosis};
\node[above =  0.75 of dscan, text centered] (filter) {$\sigma_{diag=cdiff}$};
\node[above =  0.75 of agg, text centered] (rowno) {rowno};
\node[above =  0.75 of sort, text centered] (rename) {$rename(rcd)$};


\draw[->, line width= 1] (dscan)  --  (filter);
\draw[->, line width= 1] (filter)  --  (rowno);
\draw[->, line width= 1] (rowno)  --  (rename);

\node[right =  1 of rename, text centered] (distinct) {distinct};
\node[below =  0.75 of distinct, text centered] (join) {\begin{Large}$\bowtie$\end{Large}};
\node[below left =  0.6 of join, text centered] (lrcd) {rcd};
\node[below right =  0.6 of join, text centered] (rrcd) {rcd};

\draw[->, line width= 1] (join)  --  (distinct);
\draw[->, line width= 1] (lrcd)  --  (join);
\draw[->, line width= 1] (rrcd)  --  (join);

\draw[line width= 1] (-1.35,0.7) --  (-1.35,4.25) -- node [near start,fill=white] {\small Sliced on Patient ID}  (4.5, 4.25) -- (4.5, 0.7) -- cycle;

\end{tikzpicture}
        }        \caption{Recurrent {\it C.\@Diff}}
    \end{subfigure}
    ~ 
    \begin{subfigure}[t]{0.31\textwidth}
        \centering
       	\resizebox {!} {1.5in} {
        	   \begin{tikzpicture}
\node[text centered] (count) {COUNT(*)};
\node[below = 0.75 of count, text centered] (distinct) {distinct};
\node[below =  0.75 of distinct, text centered] (join) {\begin{Large}$\bowtie$\end{Large}};
\node[below left =  0.5 of join, text centered] (dfilter) {$\sigma_{diag=hd}$};
\node[below right =  0.5 of join, text centered] (mfilter) {$\sigma_{med=aspirin}$};

\node[below = 0.75 of dfilter, text centered] (dscan) {diagnosis};
\node[below=0.75 of mfilter, text centered] (mscan) {medication};

\draw[->, line width= 1] (distinct)  --  (count);
\draw[->, line width= 1] (join)  --  (distinct);

\draw[->, line width= 1] (dfilter)  --  (join);

\draw[->, line width= 1] (mfilter)  --  (join);
\draw[->, line width= 1] (dscan)  --  (dfilter);

\draw[->, line width= 1] (mscan)  --  (mfilter);

\draw[line width= 1]  (-2.5,-0.75) -- (-2.5, -4) --  (2.8, -4) -- (2.8, -0.75) -- node [near end,fill=white] {\small Sliced on Patient ID}  cycle;

\draw[line width= 1]  (-1.2,-0.5) -- (-1.2, 0.5) -- node [near start,fill=white] {\small Secure}  (1.2, 0.5) -- (1.2, -0.5) --  cycle;



\end{tikzpicture}
        }
	\caption{Aspirin Count}
    \end{subfigure}%
    \caption{Optimized PDN query execution plans for running example.}
	\vspace{-4mm}
   \label{fig:query-plans}
\end{figure*}
\subsection{Optimized Plans}

Let's put this all together by examining our optimized {\sc \sysname} plans for the queries in our running example.   We display the query trees in Figure~\ref{fig:query-plans}.  

{\it Comorbidity} starts by executing the diagnosis scan in plaintext on each host, filtering its input tuples for ones in the {\it c.\@ diff} patient registry.  Since this query's slice key is the diagnosis code, a protected attribute, its execution cannot be trivially partitioned.  Next, it performs a split aggregate, wherein each host computes a partial diagnosis count for each condition locally.   Each site feeds its partial counts into a single unit of secure computation and the parties work together to sum up their counts for overlapping diagnoses, sort them, and take the most common ones.

{\it Recurrent c.\@ diff} runs almost entirely in sliced mode.   The filter\textendash its first {\it high} operator\textendash  checks for an infection diagnosis.   It takes  on the slice key of its parent, a window aggregate partitioned on patient ID.  The window aggregate numbers the infection diagnoses of each patient, where the diagnoses are sorted on timestamp.   We then self-join this table one patient at a time to identify the ones with a recurring diagnosis, and eliminate duplicates using the same slice key.

{\it Aspirin count} begins in plaintext with scans on the medication and diagnosis tables.   Next, we filter on a protected attribute, and this step is sliced on patient ID.  We then join the two tables to identify heart disease patients who received an Aspirin prescription.  After that, we eliminate duplicate patient IDs one slice at a time.  Finally, we switch to secure mode to count up the patient IDs over all slices.





In summary, we optimize our use of secure computation in three ways.  First,  {\sc \sysname}  horizontally partitions the data over public attributes  to reduce the time and complexity of our oblivious computing.  Second, the query evaluator splits up query operators to prolong its time in a less expensive execution mode.  Lastly, the optimizer identifies tuples that do not require distributed secure computation and evaluates them within their source DBMS.

 \section{Results}
\label{sec:results}

We now verify that  {\sc \sysname} produces efficient query plans using the workload introduced in Section~\ref{sec:running-example}.  We first review our experimental design.   Next, we explore the effectiveness of  this system's heuristics-based optimizer at managing our use of SMC.    Then we examine the impact of SMC on assorted database operators.  After that, we test the scalability of {\sc \sysname} as it executes over data of increasing size.    Lastly, we reveal the performance profile of this system in comparison to a hypothetical federated database where the parties trust one another.    

\subsection{Experimental Setup}
\label{sec:experiment-config}

We evaluate {\sc \sysname} on medical data from two Chicago-area hospitals in the HealthLNK data repository~\cite{healthlnkirb} over one year of data.  This dataset has 500,000 patient records, or 15 GB of data.  To simplify our experiments, we use a public patient registry for common diseases that maintains a list of anonymized patient identifiers associated with these conditions.  We filter our query inputs using this registry.


{\sc \sysname}'s query executor is built atop PostgreSQL 9.5 running on Ubuntu Linux. We evaluated our two-party prototype on 8 servers running in pairs.  The servers each have 64 GB of memory, 7200 RPM NL-SAS hard drives, and are on a dedicated 10 Gb/s network. Our results report the average of three runs per experiment.  Unless otherwise specified, the results show the end-to-end  runtime of a query, including its plaintext and secure execution.  All figures display their runtimes on a logarithmic scale.  

 \cut{Our plaintext distributed query evaluation is implemented using PostgreSQL's foreign data wrapper extension.  Thus all computation that does not occur in SMC happens directly in the database, minimizing overhead from data movement  and our performance benefits from standard query optimization (e.g., selecting the best join algorithm).}

\begin{figure} [t]
\centering
\includegraphics[width=0.5\textwidth]{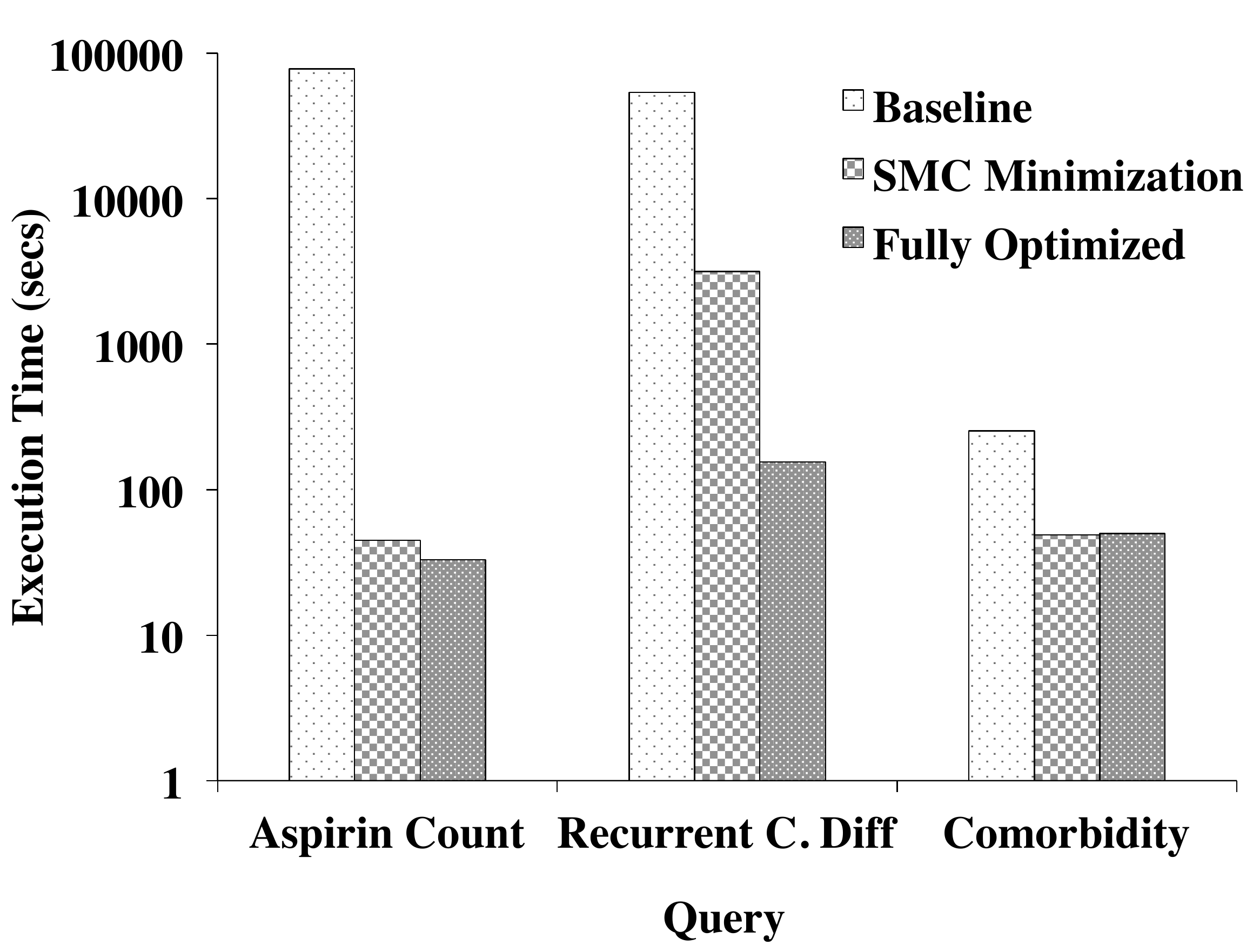} 
\caption{System performance on sampled data.}
\label{fig:end-to-end}
\vspace{-5mm} 
\end{figure}

\subsection{Secure Query Optimization}

We now examine the role of {\sc \sysname}'s optimizations in making PDN queries run with efficiency and scalability.  The results in this section are from query executions over samples of HealthLNK data with 50 tuples per data provider.  The samples were taken uniformly at random, with the restriction that each sample has at least one distributed slice partition so that the query uses secure computation.  

We evaluate the optimizer's heuristics with three tests.  First we use a baseline of fully secure execution with  no optimizations.  This test is analogous to a query execution where all of the attributes in a PDN's schema are protected.   The baseline has the same configuration as the results in Section~\ref{sec:handcoded}.   The second approach, SMC minimization, evaluates optimizations that reduce the subtree of a query's plan that is executed securely and the data processed therein.  For {\it comorbidity}, this tests split operators.   In {\it aspirin count} and {\it recurrent c.\@ diff}, it showcases the secure semi-join.   Lastly, we measure the system's performance when fully optimized.  These results show the system performance with the previous optimizations plus sliced execution.  

Figure~\ref{fig:end-to-end} displays the runtime for each query end-to-end.  It is clear that our baseline execution is very slow, even for modest data sizes.  Leveraging the PDN's security policy is important for efficient query evaluation in this setting.   The SMC minimization techniques substantially improve the system's performance for all queries.   With the split operator evaluation, {\it comorbidity} runs 5X faster than the baseline.  This query has no sliced operators, so it realizes no additional speedup in the fully optimized test.

We see the most dramatic performance improvement from {\it aspirin count}.  The secure semi-join heuristic substantially reduces the number of tuples processed in this query and it has a speedup of 1700X over the baseline.   This query benefits less from sliced execution because its tuples are not evenly distributed over the slice partitions.  This skewed distribution reduces the effectiveness of slicing because its largest partitions are comparable in size to the single partition run in the SMC minimization experiment.

In contrast, {\it recurrent c.\@ diff}'s SMC minimization has a less pronounced improvement of 17X over the baseline.  This query has fewer opportunities for optimization because its plan has a simpler tree of operators in comparison to {\it aspirin count}.   On the other hand, {\it recurrent c.\@ diff} responds better to slicing owing to its tuples having an even distribution over the query's partition keys.  Its fully-optimized version is nearly 350X faster than the baseline.

\begin{figure} [t]
\centering
\includegraphics[width=0.5\textwidth]{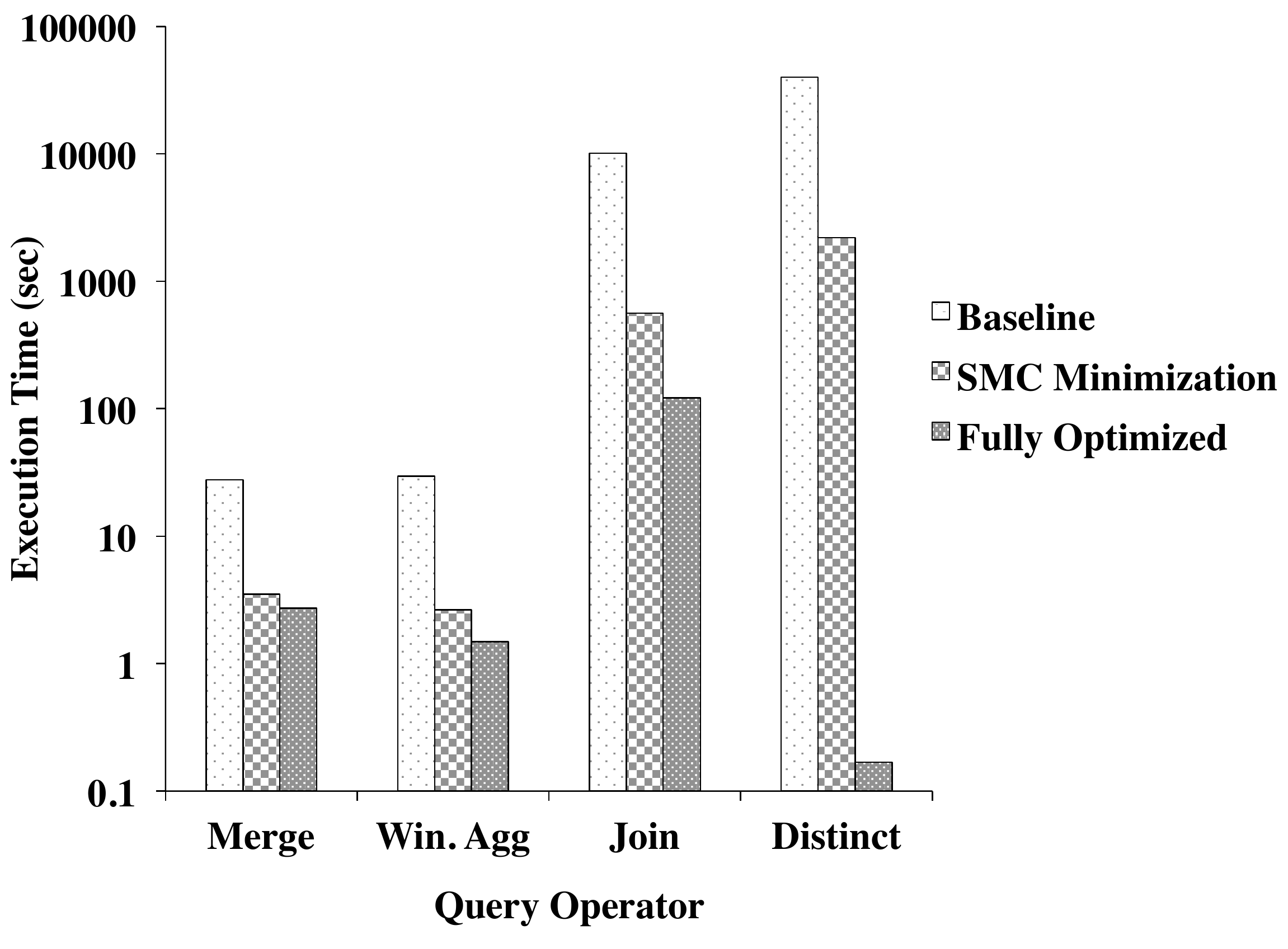} 
\caption{Optimized performance of {\it recurrent c.\@ diff}.}
\label{fig:cdiff-sliced}
\vspace{-5mm} 
\end{figure}

Let's drill down to how {\sc \sysname}'s optimizations improve the performance of {\it recurrent c.\@ diff}.  We show the breakdown of this query's runtime on each of its secure operators in Figure~\ref{fig:cdiff-sliced}.  We see a gradual performance improvement as we add more optimizations.   The merge and window aggregate operators both make simple passes over the data, and their oblivious runtimes are proportional to the size of their inputs.  The unoptimized join is costly because it computes the cross product of its input relations.   The {\tt DISTINCT} operator has a substantial SMC-induced performance penalty owing to its large input from the exhaustive join.   

Secure semi-join noticeably improves the performance of each operator.  By reducing the cardinality of the secure input data, the first two operators run much faster.  This is primarily owing to the reduced I/O costs associated  its oblivious memory accesses on smaller arrays of data.  The join shows a reduction in runtime due to its performing fewer tuple comparisons. Likewise, the final distinct operator has a strong performance improvement over the baseline owing to its smaller input data.  

We see that slicing bolsters the performance of the {\it recurrent c.\@ diff} query.  Building multiple small oblivious tuple arrays speeds up our data ingest.  Window aggregate enjoys even greater gains since slicing partitions the data by its group-by clause\textendash further simplifying its secure code.   The join also becomes faster because it is computing the cross product over fewer tuples.  The cost of finding the distinct patient IDs goes to less than a second because each slice partition simply checks if it has a non-empty array  of tuples.

\begin{figure} [t]
\centering
\includegraphics[width=0.46\textwidth]{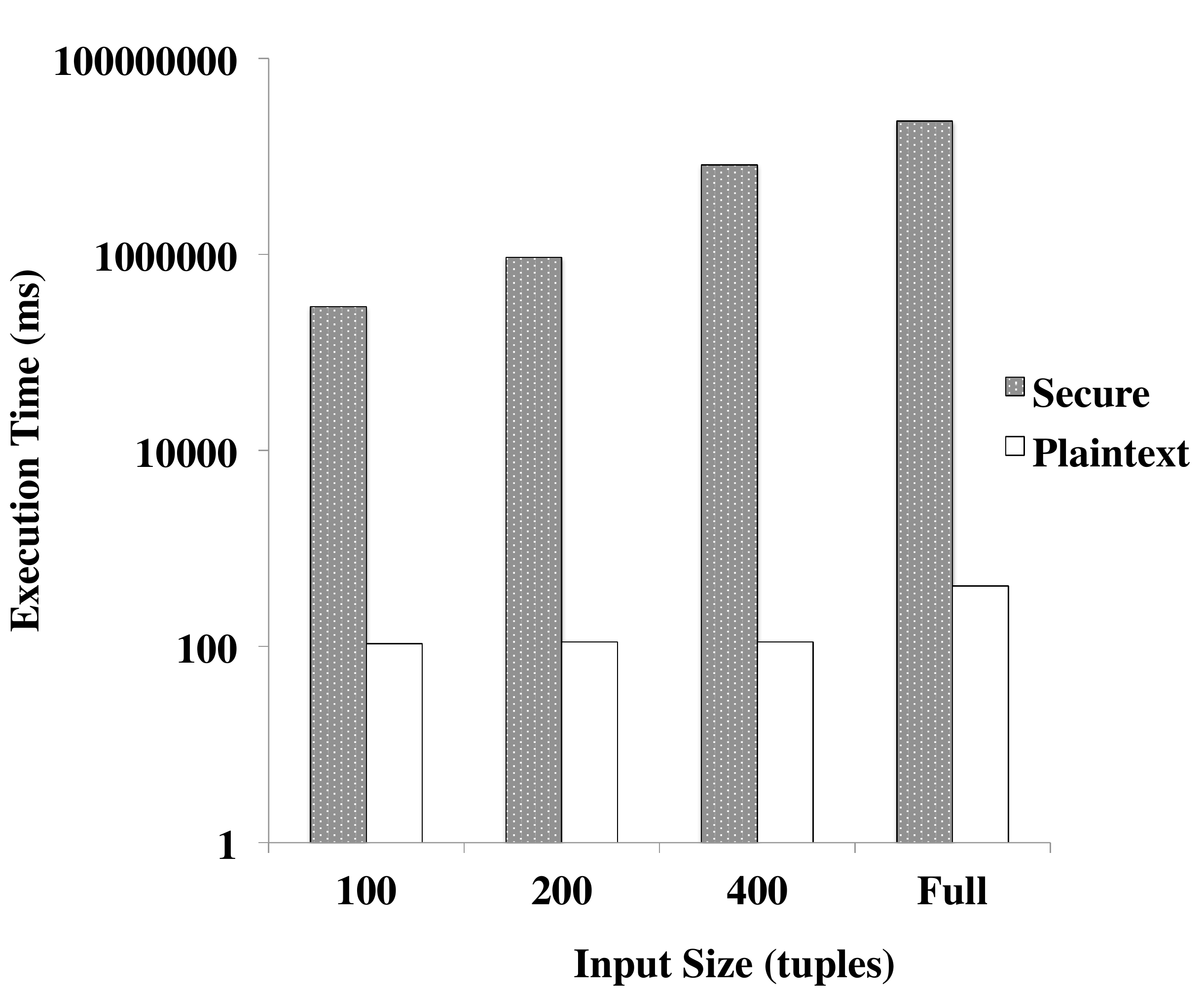}  
\caption{Runtime of {\it comorbidity} on increasing data sizes.}
\label{fig:comorbidity-scale}
\vspace{-5mm}
\end{figure}

\subsection{System Scale Up}
We now test the performance of {\sc \sysname} as it scales to larger input data.   For this experiment, we ran the {\em comorbidity} query.  We artificially limited the size of the SMC input\textendash the partial counts of comorbid conditions in the {\it c.\@ diff} cohort\textendash to 100, 200, and 400 tuples.  We ran also ran this test on the full dataset; it had nearly 800 diagnosis codes with each party supplying around 650 partial counts.  We report runtimes for this experiment in milliseconds.  

Figure~\ref{fig:comorbidity-scale} shows our scale up results.   
The plaintext execution time of this query grows slowly as we scale up.  The dominant cost of plaintext {\it comorbidity} is in local computing, wherein it scans each table, filters for the cohort, and computes the local partial count.   

In contrast, {\it comorbidity}'s secure execution shows a substantial increase in duration as the framework computes over more data.  This is primarily caused by increasing costs associated with larger oblivious tuple arrays instances for storing and accessing the data.  Because the secure array shuffles the data every time we access an element, its cost increases in proportional to its input size.  In addition, the query execution takes longer because its secure operators perform more comparisons among the inputs to sum up  partial counts.

The smallest scale run in this analysis has a slowdown of 2,700X in comparison to its plaintext execution.   With an input size of 400 tuples, the query's duration jumps to nearly 74,000X its conventionally executed counterpart.  When we run over the entire SMC input, our runtime goes to 6.5 hours or 56,000X the baseline.

\begin{figure} [t]
\centering
\includegraphics[width=0.45\textwidth]{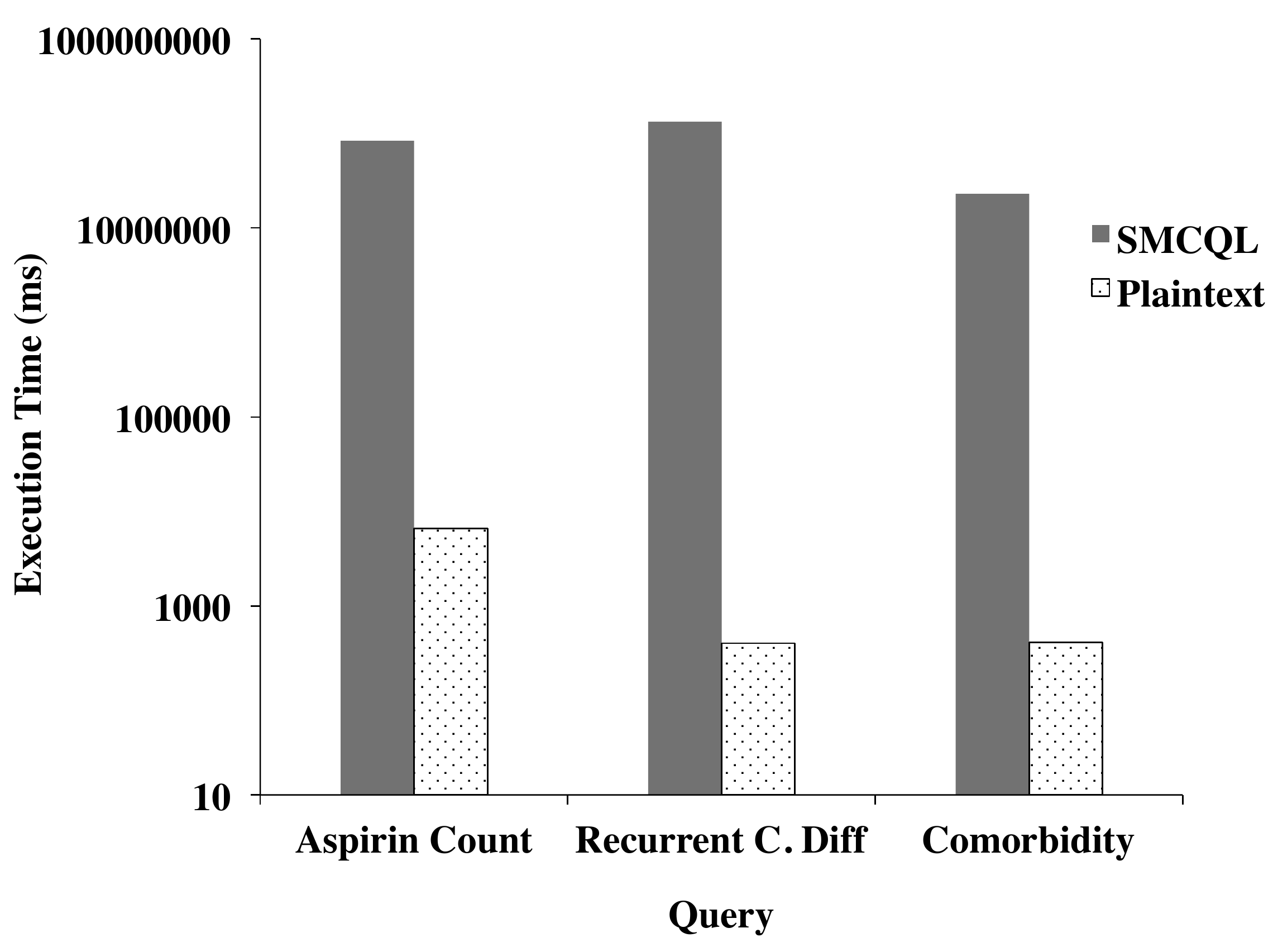}
\caption{Runtime on full year of HealthLNK records.}
\label{fig:overall}
\vspace{-3mm}
\end{figure}

Zooming out to look at the system's end-to-end performance, we now appraise the effects of its entire query optimization system.  For this experiment, we used a full year of health records.   Our results are shown in Figure~\ref{fig:overall}.  These results are a scaled-up evaluation of the fully optimized test in Figure~\ref{fig:end-to-end}.

All of these queries run 5--6 orders of magnitude more slowly than their plaintext counterparts.   {\sc \sysname} executes {\it recurrent c.\@ diff} in 37 hours, {\em aspirin count} in 23 hours, and {\em comorbidity} in 6 hours.  These queries are accessing a collective 42 million diagnoses and 23 million medication records.  By using fine-grained partitions of the underlying dataset, the SQL operators scale reasonably to large volumes of data.  

We see noticeably higher runtimes for {\em aspirin count} and {\it recurrent c.\@ diff}. Both of these take upwards of a single day to run. The primary source of the slowdown arises from their  {\tt join} operators that 
have hundreds of input tuples  and they perform $O(n^2)$ comparisons for oblivious evaluation.  More work is needed to manage this challenge of securing large in-memory data structures.

One potential solution for scaling {\sc \sysname} to large datasets is to partition the work into smaller units and use the honest broker to assemble the results.  This approach would require an analytical cost model to identify how finely to partition the work based on the rate at which the honest broker can accept and assemble the final results.  It would also require the planner to analyze the cardinality of intermediate results to ensure that the honest broker does not receive unauthorized access to protected or private attributes.  In our running example, this would mean ensuring that intermediate results received by the honest broker were  $k$-anonymous in their protected attributes and had no private data.

In summary, our results demonstrate that {\sc \sysname} provides practical secure query evaluation over SQL for mutually distrustful parties.  We found that joins are the most costly oblivious operator and that their large output cardinalities impact the execution time of their parent operators.   Slicing reduces the slowdown associated with secure query execution, and it is most effective for datasets where the tuples are evenly partitioned over the slice key space.  Our results show that query runtime grows rapidly as the system scales up to larger datasets.  

  \cut{In conducting These results expose numerous opportunities for additional PDN query optimization, especially in selecting SMC protocols, beyond our blanket use of garbled circuits and ORAM.  We also conjecture about opportunities for parallelism in {\sc \sysname}'s fine-grained query execution plans.  Further gains may be possible by combining loops in the secure operator pipeline because this rewrite would reduce the number of times we read and write to ORAM.}

\vspace{-2mm}
\section{Related Work}

Several approaches to secure query processing exist in the literature.  Most of this work is designed for outsourced computation~\cite{Laur2013,Laur2011,Popa2011}.  In contrast, a PDN keeps query processing in the database as much as possible and performs SMC within the hosts from which the data originated.   This enables data providers who do not trust the cloud (or some other third party) to share information.   

CryptDB~\cite{Popa2011} stores its data on a remote server using homomorphic encryption.  Also, in~\cite{aggarwal2005two}, the authors proposed outsourcing database query evaluation by storing the entire database as secret shares spread over two or more cloud providers.   The work of ~\cite{Laur2013,Laur2011} computes database filters and joins using SMC that is run on trusted third-party servers.  Sharemind~\cite{Bogdanov2016} supports federated databases, but does not allow for arbitrary SQL queries and uses encrypted cloud storage for their data.  Because {\sc \sysname} runs SMC locally on the hosts that provide the data, we make our queries faster by minimizing our use of SMC.  This is not possible in outsourced systems since they do not have access to the unencrypted input data.

 Chow, et al.~\cite{chow2009two}  proposed outsourced SMC for SQL queries over data from multiple source systems.  
\cut{ who then sends it to a randomizer that tells the databases how to encode intermediate query results.  The databases send these encoded results to a compute engine that evaluates their obfuscated tuples and sends the combined results to the front end.   This work assumes that the compute engine and randomizer cannot communicate.  }  Our work removes the reliance on trusted third parties by performing the secure computation within the hosts of each data provider.  In addition, the existing work did not support arbitrary tuple comparisons (e.g., $\leq$, $\geq$)  and it leaked information for nested queries.  {\sc \sysname} has both of these features.  

There is a plethora of domain-specific programming languages for working with SMC, including ObliVM~\cite{Liu2015}, VMCrypt~\cite{malka2011vmcrypt}, TASTY~\cite{henecka2010tasty}, and FAIRPLAY~\cite{malkhi2004fairplay}.  They all generate secure code for procedural programs.   These systems rely on the user explicitly specifying how to manage and compare the data from each party.  In contrast, {\sc \sysname} seamlessly injects SMC into an existing declarative language, SQL.

Kerschbaum~\cite{kerschbaum2011automatically,Kerschbaum2013} broke ground by optimizing the use of SMC in imperative programs using a secure type flow system. We extend their approach with a security type system for SQL.   This existing work operated in the context of all parties learning the plaintext output of secure computation.  The authors use this information to deduce the intermediate states of a SMC program that are safe to reveal to the data providers. Rastogi~\cite{rastogi2013knowledge} generalizes and expands upon this work using a knowledge inference approach.   In our research, PDN members do not have access to the output of secure computation.  The optimization techniques used here are more conservative than their predecessors in how they handle intermediate results.


\vspace{-2mm}
\section{Conclusions and Future Work}
In this work, we introduce the private data network (PDN), a novel generalization of federated database systems for mutually distrustfully parties.  We propose {\sc \sysname}, a framework for translating SQL queries into secure multiparty computation primitives for  evaluating PDN queries.  We introduce several strategies for optimizing PDN query plans, including partitioning data before securely computing on it  and partially evaluating database operators in plaintext.   

Our results demonstrate that by partitioning query processing at a fine granularity, we offer usable  performance for complex PDN workloads.  Our PDN queries, that are based on a real medical use case and evaluated on de-identified medical records, complete within a reasonable time even in the presence of large datasets.    We designed this system to be practical.  To this end, we are preparing {\sc \sysname} for an open source release.   In addition, we are collaborating with stakeholders in a clinical data research network to start testing it in the field.

There are numerous opportunities for future research on PDNs.  We are investigating how to generalize {\sc \sysname} to three or more parties.  Scaling out to more parties requires changes to our SMC protocols, as well additional algorithms for assigning work to PDN nodes.  PDNs with a large number of parties also introduce opportunities for parallelizing their plans.   Another  future research direction is to identify automatic SQL rewrite rules that further delay our entry into SMC by reordering commutative database operators.   Optimizing the SMC primitives and protocols used in a PDN plan is another future avenue of inquiry.    There are many choices for how to implement the secure computation within a PDN, including randomized key protocols~\cite{chow2009two} and garbled ORAMs~\cite{gentry2014garbled}.  In addition, there are garbled circuit protocols optimized for reduced CPU time, minimizing network bandwidth, and for scaling to a large number of parties.   Lastly, extensive research exists on SMC protocols for specific algorithms, such as linear regression and matrix multiplication, but we are aware of no work on improving the performance of secure SQL operators.

\section{Acknowledgments}
We thank Katie Jackson and Jess Joseph Behrens for their guidance and assistance with CAPriCORN and HealthLNK data.  We appreciate the HealthLNK team for sharing de-identified electronic health record data for this study.  We are grateful to Ben Slivka and Mike Stonebraker for their feedback on early drafts of this work.

\balance

\bibliographystyle{abbrv}
{\scriptsize \bibliography{smcql} }


\end{document}